\documentclass[prb,twocolumn,superscriptaddress]{revtex4}
\usepackage{graphicx}
\begin{document}

\title[Vibrational properties of  LiBC]{
Vibrational properties of hexagonal LiBC: Infrared and Raman
spectroscopy}

\date{\today}

\author{J. Hlinka}
\author{ V. \v{Z}elezn\'y}
\author{ I. Gregora}
\author{ J. Pokorn\'y}
\affiliation{Institute of Physics ASCR, Praha, Czech Republic}
\author{A. M. Fogg}
\author{J. B. Claridge}
\author{G. R. Darling}
\author{M. J. Rosseinsky}
\affiliation{ Department of Chemistry, University of Liverpool,
 Liverpool L69 3BX, UK}

\begin{abstract}
The paper presents infrared reflectivity and micro-Raman
scattering spectra of LiBC powder pellets. The experiment allowed
assignment of frequencies of all infrared and Raman active zone
center modes: E$_{1u}$(LO) at 1262~cm$^{-1}$ and 381~cm$^{-1}$,
E$_{2g}$ at 1172~cm$^{-1}$ and 174~cm$^{-1}$ and A$_{2u}$(LO) at
825~cm$^{-1}$ and 545~cm$^{-1}$. Results are compared with
available ab-initio calculations; prediction of large Born
effective charges on the nodes of B-C graphene sheets is
confirmed.
\end{abstract}


\maketitle

LiBC is a layered boron carbide consisting of alternating
graphene-like (BC)$^-$ sheets separated by intercalated Li$^+$
ions. It normally crystallizes with a hexagonal structure of $\rm
D_{6h}^4$ ($\rm P6_3/mmc$) space group symmetry with Li, B and C
atoms in 2a, 2c and 2d Wyckoff positions,
respectively\cite{Wor95}. The structure is very close to that of
the recently discovered unconventional superconductor
MgB$_2$\cite{nature1}. Electronic band structure of both materials
is also quite similar, except for that LiBC is an insulator with
completely filled 2p-$\sigma$ graphene bands. Since the
deformation potential due to the $E_{2g}$ zone center bond
stretching mode is in LiBC even higher than in
MgB$_2$\cite{Ros02}, it was predicted that the hole-doped LiBC
could show superconductivity with $T_{\rm c}$ of order of 80\,K.
Several groups\cite{Cava,Souptel,Zhao,FoggPRB,FoggChC,Ren03} tried
different methods to achieve superconductivity in Li deficient
samples, but none of these attempts were successful. The reason of
the failure (or failure of the prediction) has not yet been
elucidated. In any case, comparative LiBC {\it vs} MgB$_2$ studies
are desirable for detailed understanding of the MgB$_2$-type
superconductivity.

Vibrational properties of LiBC were thoroughly studied by
ab-initio methods\cite{ARSP,ASRP,Kwan,Dew,Ren03}, but due to the
lack of large single crystals, the desirable experimental
information is quite limited.\cite{Artem,Ren03,Bha,ourLiBC1}
Group-theoretical analysis predicts ten zone-center optic lattice
modes: a pair of Raman active $E_{2g}$ modes (B-C bond stretching
mode and B-C layers sliding mode); 2$E_{1u}$
 (B-C bond stretching mode and B-C
layer {\it vs} Li layer sliding mode) and $2A_{2u}$
 (B-C layer puckering mode and  B-C
layer against Li layer beating mode) infrared active modes; and
2$B_{2g}+E_{2u}+B_{1u}$ optically silent modes. In this paper, we
present results of a systematic room-temperature infrared and
Raman spectroscopic study on polycrystalline LiBC pellets, which
provides a complete spectrum of zone center optically active modes
in LiBC ( 2$E_{1u}+ 2E_{2g}+ 2A_{2u}$ species.)

Let us briefly review the previous experimental investigations of
phonons in LiBC by infrared, Raman and inelastic neutron
scattering spectroscopy on microcrystals and powder samples.
Inelastic neutron scattering has shown weighted phonon density of
states extending up to about 1300\,cm$^{-1}$, with three
pronounced bands in the range 350--450\,cm$^{-1}$,
700--850\,cm$^{-1}$ and 1000--1250\,cm$^{-1}$, corresponding to
external, puckering and stretching modes of the graphene-like
sheets, respectively (the lowest frequency band comprises also
Li-ion vibrations.) A pair of Raman active $E_{2g}$ modes was
observed\cite{ourLiBC1,Ren03,FoggPRB} near 170\,cm$^{-1}$ and
1170\,cm$^{-1}$. These modes correspond to sliding of the graphene
sheets and to the B-C bond stretching modes, respectively. In
addition, another pair of sharp and strong Raman lines, presumably
corresponding to $B_{1g}$ modes, was seen in a metastable trigonal
form of LiBC\cite{ourLiBC1}. Two of four infrared active modes
($E_{1u}$ species) should contribute to the reflectivity of
hexagonal faces. However, the infrared microscope
experiment\cite{Artem} on a micro-crystallite with a
well-developed natural hexagonal face showed a more complicated
spectrum, so that only the higher frequency $E_{1u}$ (at
1180\,cm$^{-1}$) could be reliably assigned.\cite{Artem} The other
two infrared active modes, polarized along the hexagonal axis
($A_{2u}$ species), should contribute together with $E_{1u}$ modes
to the infrared response of powder samples. Unfortunately, the
previously published\cite{Bha} reflectivity and transmission
spectra on LiBC powder are far from the expected 4-mode spectral
profile.

Samples used in this study were prepared at the University of
Liverpool. Stoichiometric LiBC was synthetized in Ta ampoules at
1773\,K under Ar atmosphere by the method described in
Refs.~\onlinecite{Wor95,FoggPRB}. The golden polycrystalline
powder, handled under inert atmosphere, was characterized by
laboratory x-ray diffraction test proving a single LiBC phase
 with lattice parameters $a=2.75$\,\AA~and
$c=7.05$\,\AA. On a closer inspection, small systematic shoulders
on the Bragg reflections were found, indicating\cite{FoggChC} a
small amount of Li deficient phase with composition of about
Li$_{0.95}$BC ($a=2.74$\,\AA, $ c=7.07$\,\AA).
 The powder was then isostatically pressed to
form 0.65\,mm thick pellets with 8\,mm diameter. Spectroscopic
experiments were carried out in IOP ASCR in Praha within 20 hours
after opening of the sealed glass ampoules containing the pellets.

 The Raman experiments were carried out using a Renishaw Raman
microscope with 514.5~nm (2.41~eV) argon laser excitation. The
instrument allows both the direct microscope observation and
measurement of polarized Raman spectra in back scattering
configuration from a spot size down to 1-2 microns in diameter. To
minimize heating of the sample in the laser focus, the laser power
was kept below 1mW.

\begin{figure}[h]
 \hspace{0cm} \centerline{\includegraphics[width= 6cm, clip=true] {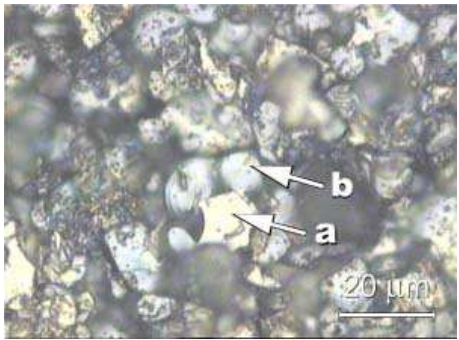}}
\caption{ (Color online) Optical microscope view of the surface of
the virgin LiBC pellet, showing "yellowish" (a) and "bluish" (b)
regions.\label{Fig1}}
\end{figure}

\begin{figure}[h]
 \hspace{0cm} \centerline{\includegraphics[width= 6cm, clip=true] {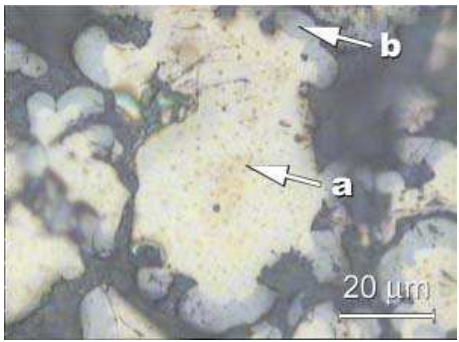}}
\caption{ (Color online) Optical microscope view of a larger grain
on the surface of the partially polished LiBC pellet.\label{Fig2}}
The "yellowish" (a) and "bluish" (b) regions correspond to
stoichiometric (LiBC) and non-stoichiometric (Li$_{0.95}$BC)
compositions, respectively.
\end{figure}

Surface of virgin pellets showed a dark golden-brown metallic
appearance at naked eye view, but optical microscope observations
revealed crystallites with bluish and yellowish faces with typical
size of order of 10 microns. The borders of bluish faces were
often rounded or kidney-shaped, while the borders of the yellowish
faces were more straight (Fig.~1). Residual area corresponded to
holes or black material without any Raman signal. After polishing
of the surface, it became apparent that the border area of larger
crystallites tends to be bluish, while the interior part is
yellowish, with a well defined boundary between the bluish and
yellowish regions (Fig.~2). Raman spectra of the bluish regions
show a pair of $E_{2g}$ modes near 160\,cm$^{-1}$ and
1184\,cm$^{-1}$ and weak, broad features d$_1$, d$_2$ and d$_3$
reminiscent of the phonon density of states bands, superposed on a
strong luminescent background (see Fig.~3). Very similar Raman
spectra were observed previously on the annealed LiBC in Ref.
\onlinecite{ourLiBC1}. In contrast, the luminescent background was
practically absent in the yellowish crystallites, and the $E_{2g}$
lines were significantly sharper and at a somewhat "repelled"
positions 174\,cm$^{-1}$ and 1172\,cm$^{-1}$ (see Fig.~3).
Furthermore, the yellowish crystallites revealed strong asymmetric
bands near 1700\,cm$^{-1}$ and 2500\,cm$^{-1}$, which strongly
reminiscent of two-phonon double resonant Raman scattering lines
in graphite.\cite{ThomsenReich} These bands, first reported in
Ref.~\onlinecite{FoggPRB}, indeed correspond well to doubled
frequency of the puckering and B-C bond stretching vibrations, and
will be investigated in more detail elsewhere. From the above
observations, we conclude that the yellowish regions correspond to
the stoichiometric LiBC, while bluish regions correspond to the
non-stoichiometric Li$_{0.95}$BC component seen by X-ray
diffraction.

\begin{figure}[h]
 \hspace{0cm} \centerline{\includegraphics[width= 8cm, clip=true] {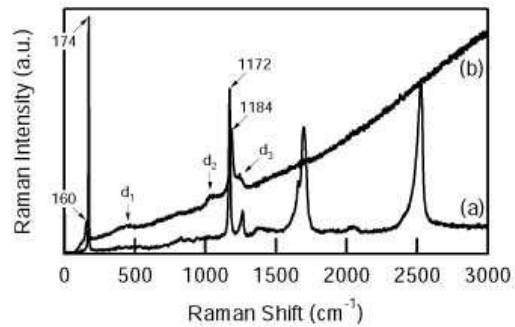}}
  \caption{ Typical unpolarized Raman spectrum taken
from yellowish (a) and bluish (b) part of the LiBC
pellet.\label{Fig3}}
\end{figure}

Finally, let us stress that none of the Raman spectra taken from
this sample showed the additional pair\cite{ourLiBC1} of sharp and
strong $B_{1g}$-like Raman lines near 546\,cm$^{-1}$ and
830\,cm$^{-1}$, so that the present sample is clearly free from
the low symmetry modification. We have observed, however, in some
of the yellowish regions, a very weak but quite sharp lines near
388\,cm$^{-1}$, 548\,cm$^{-1}$ and 828\,cm$^{-1}$, which, as will
be shown below, correspond surprisingly well to the LO frequencies
of infrared active optic modes. We speculate that these weak
features may be coupled LO phonon-plasmon modes.\cite{plasmon} In
this case, such modes should be absent in cross-polarized
geometry, which was indeed observed (see Fig.~4).

\begin{figure}[h]
 \hspace{0cm} \centerline{\includegraphics[width= 8cm, clip=true] {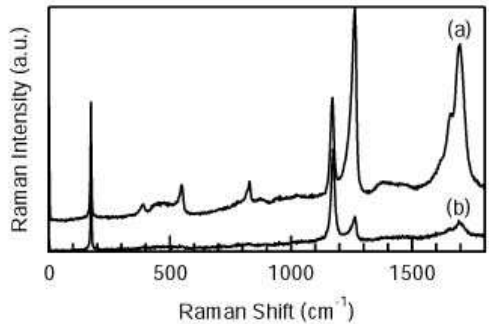}}
 \caption{ Raman spectra of a yellowish crystallite
showing three weak, normally forbidden lines close to LO
frequencies of infrared active modes. Spectra are taken in (a)
parallel and (b) perpendicular polarization conditions.
\label{Fig4}}
\end{figure}

Infrared reflectivity at near-normal incidence was measured using
a Bruker 113v spectrometer. To improve the surface quality, we
tried both dry and wet polishing using diamond paste and different
organic liquids, but we were not able to achieve a mirror-like
reflection over the entire surface of the pellet. Therefore, we
have rather measured directly the reflectivity of the as received
("virgin") surface. The absolute value of reflectivity is
calculated as a ratio of of the sample and Al mirror spectra.
After the measurement, about 300 nm of Au was evaporated on the
measured pellet surface, in order to perform an auxiliary
reflectivity measurement allowing to estimate the area of highly
reflecting microcrystalline faces arranged parallel to the
surface. Reflectivity of these surfaces was then determined as a
ratio of virgin pellet and Au-coated reflectivities.

\begin{figure}[h]
 \hspace{0cm} \centerline{\includegraphics[width= 8cm, clip=true] {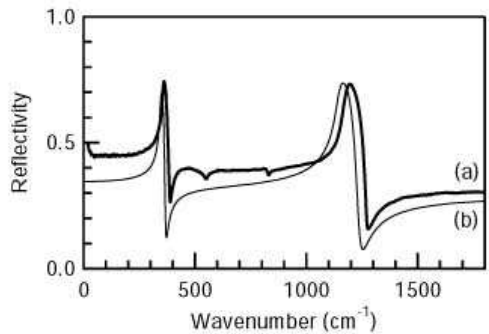}}
  \caption{ Unpolarized infrared reflectivity spectrum
of the LiBC. (a) Typical unpolarized experimental spectrum of the
investigated LiBC pellet. (b) Theoretical single-crystal c-face
reflectivity calculated from the model using ab-initio calculated
phonon parameters (details in the text).\label{Fig5}}
\end{figure}

Resulting reflectivity spectrum (Fig.~5.) shows two clear bands
corresponding to E$_{1u}$ phonon modes (with LO frequencies near
1262~cm$^{-1}$ and 381~cm$^{-1}$). This is obvious from comparison
with the single-crystal reflectivity calculated for normal
incidence c-face reflection
\begin{equation}
 R_{a}(\omega) =
\left|\frac{\sqrt{\epsilon_{a}{(\omega)}}-1
}{\sqrt{\epsilon_{a}{(\omega)}}+1 } \right|^2 ~~,
\end{equation}
using the usual damped harmonic oscillator expression for in-plane
dielectric permittivity $\epsilon_{a}(\omega)$
\begin{equation}
\frac{\epsilon_{a}(\omega)} {\epsilon_{\rm a}^{\infty}} = 1 +
 \frac{\Omega_1^2}{\omega_1^2 - \omega^2 -i\omega \Gamma_1}
+\frac{\Omega_2^2}{\omega_2^2 - \omega^2 -i\omega \Gamma_2}~~,
\end{equation}
with {\it ab-initio} calculated\cite{Kwan} parameters (electronic
permittivity $\epsilon_{ \rm a}^{\infty}= 11.24$, E$_{1u}$(TO)
frequencies $\omega_1 = 346$\,cm$^{-1}$, $\omega_2 =
1143$\,cm$^{-1}$, screened plasma mode frequencies $\Omega_1
=135$\,cm$^{-1}$, $\Omega_2 =469$\,cm$^{-1}$) and assuming a
reasonable damping $\Gamma_i = 0.03 \omega_i$ as in
Ref.~\onlinecite{Kwan}. The general agreement indicates that the
majority of microcrystalline faces on the surface are parallel to
the hexagonal plane, as could be guessed from the typical
plate-like habitus of LiBC powder grains.
 Two small additional dips near
545~cm$^{-1}$ and 825~cm$^{-1}$ are close to ab-initio frequencies
of A$_{2u}$(LO) modes, suggesting that few crystallites on the
surface have nevertheless a different orientation. While the
values of TO and LO frequencies of E$_{1u}$ modes could be easily
adjusted to match the experimental data, the overall increase of
the measured reflectance between 1500 and 200\,cm$^{-1}$ cannot be
attributed to the dielectric contribution of these phonon modes
only. This additional contribution could be
 an effect related to the powder form of the LiBC sample or due
to a metallic impurity component in the sample etc. On the other
hand, the sharp increase of the reflectivity below 50\,cm$^{-1}$
could be modeled by a Drude model with $\omega_p^2/\Gamma_p
\approx 10-20$\,cm$^{-1}$ what might be considered as intrinsic
LiBC effect compatible with dc conductivity of LiBC\cite{Souptel}.

\begin{table}[h]
\begin{center}
\begin{tabular}{l|rrr|r|r}
       mode & \multicolumn{3}{|c|}{ \it ab-initio} &
         \multicolumn{1}{|c|}{   Raman} &   \multicolumn{1}{|c}{ IR}  \\
& Ref.~\onlinecite{Ren03} & Ref.~\onlinecite{ARSP} &Ref.~\onlinecite{Kwan} & &\\
 \hline
E$_{2g}$ & 176& 171& 169 & 174&    \\
E$_{2u}$ & 301& 306& 292 &    &    \\
B$_{1g}$ & 319& 289& 299 &    &    \\
E$_{1u}
   ({\rm TO1})$ & 354& 352& 346 &    & 356\\
E$_{1u}
   ({\rm LO1})$ & 382&    & 367 &(388)& 381\\
A$_{2u}
   ({\rm TO3})$ & 457& 422& 407 &    &    \\
A$_{2u}
   ({\rm LO3})$ & 563&    & 499 &(548)& 545\\
B$_{2u}$ & 548& 540& 510 &    &    \\
 \hline
A$_{2u}
   ({\rm TO4})$ & 819& 802& 803 &    &    \\
A$_{2u}
   ({\rm LO4})$ & 840&    & 833 &(828)& 825\\
B$_{1g}$ & 843& 821& 829 &    &    \\
 \hline
E$_{1u}
   ({\rm TO2})$ &1136&1194&1143 &    &1174\\
E$_{1u}
   ({\rm LO2})$ &1231&    &1236 &    &1262\\
E$_{2g}$ &1145&1204&1153 &1172&  \\

\end{tabular}
\end{center}
\caption{Frequencies of zone center modes in LiBC (in cm$^{-1}$).
Values in brackets correspond to the weak sharp lines discussed in
the text.}
\end{table}

 Frequencies of all
measured phonon modes are compared with available {\it ab-initio}
calculations in Tab.~I. From Raman measurement, only the data from
the inner yellowish regions are shown. It is remarkable that the
LO frequencies calculated as zeros of the adjusted dielectric
permittivity coincides within 10\,cm$^{-1}$ with the LO
frequencies determined from Raman measurements. Generally, the
experimental frequencies of E$_{1u}$ and of E$_{2g}$ modes tend to
be somewhat higher than the theoretical ones.

The $E_{1u}$ mode experimental screened plasma frequencies
($\Omega_{1}$=147\,cm$^{-1}$ and $\Omega_{2}$=459\,cm$^{-1}$) are
quite close to the {\it ab-initio} calculated values 135 and
469\,cm$^{-1}$. These values can be used for evaluation of
in-plane diagonal components of Born effective
 charge tensors. Let us consider $E_{1u}$ modes polarized along the
x-axis. The eigenvector of the $j$-th mode can be defined by three
nonzero components of its mass-reduced polarization vectors
$(x_{\rm Li}(j),x_{\rm B}(j),x_{\rm C}(j) )$, $x_{\rm
Li}(j)^2+x_{\rm B}(j)^2+ x_{\rm C}(j)^2= 1$. The screened plasma
frequency of the mode ($j$) is then given by
\begin{equation}
\Omega_{j} = \left| \sum_{\kappa={\rm Li,B,C}} x_{\kappa}(j)
\Omega_{{\rm ion},\kappa}
\frac{Z^{*}_{\kappa,a}}{|Z^{*}_{\kappa,a}|} \right|~~,
\end{equation}
where
\begin{equation}
\Omega_{{\rm ion},\kappa} = \beta
\frac{Z^{*}_{\kappa,a}}{\sqrt{m_{\kappa}}}~~,
\end{equation}
is the ionic (in-plane) screened plasma frequency,
 $Z^{*}_{\kappa,a}$ is the in-plane diagonal components of Born effective
 charge tensor of ion $\kappa$ and $m_{\kappa}$ is its relative mass.
The common factor
\begin{equation}
\beta =\frac{e}{\sqrt{ m_{\rm u} \epsilon_0 \epsilon_{\rm
a}^{\infty} V_0}}
\end{equation}
includes elementary charge $e$, atomic mass unit $m_{\rm u}$,
permittivity of vacuum $\epsilon_0$, volume of primitive unit cell
$V_0$ and relative in-plane electronic
 permittivity $\epsilon_{\rm a}^{\infty}$.

Assuming that the lower frequency $E_{1u}({\rm TO1})$ mode
involves purely rigid motion of of graphene sheets, eigenvectors
$x_{\kappa}(j)$ of TA, TO1 and TO2 $E_{1u}$ modes are given by
columns of the matrix $T_{\kappa j} = x_{\kappa}(j)$
\begin{equation}
T=\left(
\begin{array}{ccc}
\sqrt{m_{\rm Li}} & -\sqrt{m_{\rm BC}} & 0\\
\sqrt{m_{\rm B}} & \sqrt{\frac{m_{\rm Li} m_{\rm B}}{m_{\rm BC}}}&
- \sqrt{\frac{m_{\rm C} M} {m_{\rm BC}}}\\
\sqrt{m_{\rm C}} & \sqrt{\frac{m_{\rm Li} m_{\rm B}}{m_{\rm BC}}}&
 \sqrt{\frac{m_{\rm B} M} {m_{\rm BC}}}\\
\end{array}
\right)\frac{1}{ \sqrt{M}} ~~,
\end{equation}
where $m_{\rm BC}=m_{\rm C}+m_{\rm C}$ and $M=m_{\rm BC}+m_{\rm
Li}$. Using the experimental values of screened plasma frequencies
of TO1 and TO2 modes ($\Omega_{1}$ and $\Omega_{2}$) and for
$Z^*_{\rm Li, a} > 0$, $Z^*_{\rm C, a} < 0$, $Z^*_{\rm Li, a}
+Z^*_{ \rm B,a} + Z^*_{\rm C, a} =0 $, eqns. (3 and (6)yield
unique solution $ \Omega_{\rm ion,Li}= 129\,{\rm cm}^{-1}$,
$\Omega_{\rm ion,B}=284 \,{\rm cm}^{-1}$, $\Omega_{\rm ion,C}=
-367\,{\rm cm}^{-1}$. The corresponding in-plane diagonal
components of Born effective
 charge tensors directly follows from eqs.(4) and (5), giving
(for $V_0 =23$\AA$^3$ and $\epsilon_a^{\infty}=11.24$)
\begin{equation}
Z^*_{\rm Li,a}= 0.78~,~ Z^*_{\rm B,a}= 2.15~,~ Z^*_{\rm C,a}=
-2.93~.
\end{equation}
These values are indeed close to {\it ab-initio} values\cite{Kwan}
0.81, 2.37 and -3.17.

In conclusion, although the present experiment cannot substitute
single crystal measurements, we were able to observe all optically
active modes of LiBC and estimate their frequencies. The screened
plasma frequencies frequencies of $E_{1u}$ modes were used to
determine in-plane diagonal components of Born effective charge
tensors at Li, B and C ions. The obtained results are in a perfect
agreement with first-principle predictions for vibrational
properties of stoichiometric LiBC.

\vspace{1 cm}

\begin{acknowledgements}
The work has been supported by Czech grant project AVOZ 1-010-914.
\end{acknowledgements}

\end{document}